\documentstyle[12pt,a4wide]{article}
\begin{document}
\newcommand{\be} {\begin{equation}}
\newcommand{\ee} {\end{equation}}
\newcommand{\ba} {\begin{eqnarray}}
\newcommand{\ea} {\end{eqnarray}}
\newcommand{\e} \epsilon
\newcommand{\la} \lambda
\newcommand{\La} \Lambda 

\author{B. Boisseau\thanks{E-mail :
boisseau@celfi.phys.univ-tours.fr}\\
\small Laboratoire de Math\'ematiques et Physique Th\'eorique\\
\small CNRS/UPRES-A 6083, Universit\'e Fran\c{c}ois Rabelais\\
\small Facult\'e des Sciences et Techniques\\
\small Parc de Grandmont 37200 TOURS, France}
\title{\bf Vortex in a relativistic perfect isentropic fluid and Nambu Goto 
dynamics}
\date{}
\maketitle

\begin{abstract}

By a weak deformation of the cylindrical symmetry of a potential vortex in a relativistic perfect isentropic fluid, we study  the possible dynamics of the central line of this vortex. In ``stiff''material the Nambu-Goto equations are obtained.

\end{abstract}

\section{Introduction}

These last years, the motion of the relativistic vortices has been in general studied  in the context of a model based on nonlinear wave equations of a complex scalar field. These studies enlight the theoretical similarities between superfluid vortices and global cosmic strings \cite{BY5,BY6,BY8,BY9,BY10}.

In this work, we take a purely hydrodynamics point of view  The purpose is to find the relativistic dynamics of the central line of a potential vortex in an isentropic perfect fluid when it deviates weakly of the stationarity and of the cylindrical symmetry.

 We  suppose that the core of this vortex
is thin and that its central line has a small deviation from the straight configuration; so that locally, in the neighbourhood of the core (which is empty), the flow must approach the cylindrical symmetry. Therefore, expanding the
 equations of motion in function of small parameters caracteristic of the
 curvature and the thickness of the core, the lower order which is the equation of the straight vortex, is automatically verified. The following order gives
 approximatly the dynamical equation of the central line. This method has
been transposed from a study of the dynamics of a  self-gravitating thin
cosmic string \cite{BCL}.

The plan of the paper is the following. In section 2, we shall review
the essential ingredients of a perfect fluid and give the formulation of Lichnerowicz \cite{lich} in which it is easy to write the stationnary cylindrical solutions of an isentropic fluid  in Minkowski spacetime .   In
section 3, by perturbating  the cylindrical symmetry, we obtain the dynamical equation of the central line. The interesting result is that in stiff matter it is reduced to the Nambu-Goto dynamics. In section 4 we discuss the solutions 
and finally  we end with a conclusion.

\section{Potential vortex in a perfect isentropic fluid}

Let us consider a perfect fluid \cite{landau}. The tensor energy-momentum is
\be
\label{1}
T_{\mu \nu}=(\rho+p)u_\mu u_\nu+pg_{\mu \nu}
\ee
where $\rho$ is the proper energy density, $p$ the pression and $u_\mu$ the unitary 4-velocity.The numerical current is
\be
\label{2}
j_\mu=nu_\mu
\ee
where n is the proper density of particle.

The motion of the fluid is governed by the equations of conservation
\be
\label{3}
\nabla_\mu T^{\mu \nu}=0
\ee
\be
\label{4}
\nabla_\mu j^\mu=0
\ee 

The quantities $\frac{1}{n}$ and $\frac{\rho}{n}$ are respectively the volume and the proper energy per particle. The enthalpy per particle is
\be
\label{5}
h=\frac{\rho+p}{n} .
\ee
The entropy $s$ and the temperature $T$ are defined in the thermodynamic relation
\be
\label{6}
dh=Tds+\frac{1}{n}dp 
\ee
so that $h$, $T$ and $\frac{1}{n}$ can be considered as functions of $s$ and 
$p$. In particular $h=h(s,p),~n=n(s,p)$. It will be usefull to inverse  the 
first relation and have
\be
\label{7}
p=p(s,h)\quad ,\quad n=n(s,h)
\ee
Contracting the equation of motion (\ref{3}) with $u^\nu$ gives
\be
\label{8}
\nabla _{\mu} (nhu^{\mu})-u^{\mu} \partial _{\mu}p=0.
\ee
Using (\ref{8}), the equation (\ref{3}) can be written
\be
\label{9}
nhu^{\mu}\nabla _{\mu}u_\nu+(g^\mu_\nu+u^\mu u_\nu)\partial _\mu p=0
\ee
Hence with (\ref{6}) this equation is finally written
\be
\label{10}
u^{\mu}\nabla _{\mu}u_\nu+(g^\mu_\nu+u^\mu u_\nu)(\frac{\partial _\mu h}{h}
-\frac{T\partial _\mu s}{h})=0
\ee
and the equation (\ref{8}) with (\ref{6}) gives
\be
\label{11}
u^\mu \partial _\mu s=0.
\ee
The entropy is conserved along a streamline.

The following treatment is given by Lichnerowicz \cite{lich}. We suppose that
the motion of the fluid is isentropic that is to say
\be
\label{12}
ds=0
\ee
The equation (\ref{6}) is reduced to
\be
\label{13}
dh=\frac{1}{n}dp.
\ee
In the relations (\ref{7}) we can forget the dependence in s and have
$p=p(h),~n=n(h)$.

The equation (\ref{10}) becomes
\be
\label{14}
u^{\mu}\nabla _{\mu}u_\nu+(g^\mu_\nu+u^\mu u_\nu)(\frac{\partial _\mu h}{h})=0
\ee
Let us introduce the quadrivector $C$ 
\be
\label{15}
C_\mu=hu_\mu
\ee
and its exterior differential $\Omega=dC$ which is named vorticity. Its components are
\be
\label{16}
\Omega_{\mu \nu}=\nabla _\mu C_\nu-\nabla _\nu C_\mu
\ee
The equation (\ref{14}) can be written
\be
\label{17}
C^\mu \Omega_{\mu \nu}=0
\ee
or $i_C \Omega=0$.

The Lie derivative of $\Omega$ along $C$ 
\be
\label{18}
{\cal L}_C \Omega=(di_C+i_C d)\Omega=0
\ee
from the preceeding equation and $d\Omega=0$. It is the Helmholtz theorem for an isentropic fluid: The vorticity is conserved in the motion.

The motion is called irrotational if its vorticity is null $\Omega=0$, that is
 $dC=0$. If $\Omega=0$ on spacelike hypersurface, $\Omega$ is null
everywhere by the Helmholtz equation and the motion is irrotationnal. In this case the equations are
\be
\label{19}
\nabla _\mu C_\nu-\nabla _\nu C_\mu=0
\ee
\be
\label{20}
\nabla _\mu j^\mu=0
\ee
with
\be
\label{21}
j_\mu=\frac{n(h)}{h}C_\mu \quad ,\quad h^2=-C^\nu C_\nu.
\ee

Let us note that the pression is $p=\int n(h)dh$ and that the velocity of
sound $v=\sqrt{\frac{dp}{d\rho}}$ is given by
\be
\label{22}
\frac{1}{v^2}=\frac{n'(h)h}{n(h)}
\ee

The invariance of the flow $C_\mu$ in the transformation generated by a Killing
vector $\xi$ is given by ${\cal L}_\xi C=0$ .This condition ensures also
the invariance of $j_\mu$ since $\frac{n(h)}{h}$ is only dependent of
$h=\sqrt{-C_\mu C^\mu}$ . If the fluid is irrotational we have
\be
\label{23}
{\cal L}_\xi C=(di_\xi +i_\xi d)C=d(i_\xi C)=0
\ee
hence
\be
\label{24}
C_\mu \xi^\mu=constant
\ee

In a Minkowski space equiped with cylindrical coordinates
\be
\label{25}
ds^2=-dt^2+dz^2+dr^2+r^2d\phi^2,
\ee 
let us consider a stationnary cylindrical solution of the equations
(\ref{19}) and (\ref{20}). We call this motion a potential cylindrical vortex.
 The symmmetries are generated by three Killing vectors
$\xi=\frac{\partial}{\partial t}, \frac{\partial}{\partial z}, \frac{\partial}
{\partial \phi}$, so that
\be
\label{26}
C_\mu=(-E , L , 0 , M)
\ee
where $E$, $L$, $M$ are the three contants of motion associated with the three
vectors of Killing. In (\ref{26}) we have also supposed that there is no source or sink in the core of the vortex. In consequence the radial component is null
 which ensures the equation (\ref{20}). There is yet an indetermination on
 $n(h)$ which can be resolved by an equation of state.
By a Lorentz transformation along $z$ we can choose $L=0$ and have a circular
flow. The scalar quantity $h$ is calculated:
\be
\label{27}
h^2=-C^\mu C_\mu=E^2-L^2-\frac{M^2}{r^2}.
\ee
We see from (\ref{27}) that $C_\mu$ becomes spacelike in the neighbourhood of
$r=0$ which is the central line of the vortex. So a central region, named the core of the vortex, must be empty which can be ensured by the equation of state of the fluid. We shall consider two different situations: The first is given by the familiar polytropic equation of state \cite{sha} $p=kn^\gamma$ which can be rewritten using (\ref{5}) and (\ref{13}) $n=\frac{\gamma-1}{k\gamma}(h-m)^{\frac{1}{\gamma-1}}$. The quantity $m$ is a constant of integration interpreted as the masse of a particle of the fluid. ($\gamma$ is equal to $\frac{5}{3}$ or $\frac{4}{3}$ following that the fluid is non relativistic or ultrarelativistic). The second is the less usual stiff material where the the velocity of sound is equal to the velocity of light($v=1$). Integrating (\ref{22}) gives $n=kh$. The radius
$r_0$ of the core is defined when the density $n$ of the particules becomes 
null starting from the exterior. For the polytropic equation it is given by
$h(r_0)=m$ and for the stiff material by $h(r_0)=0$.

\section{Perturbation of the cylindrical vortex}

We consider now a little deformation of the relativistic cylindrical vortex
described above. Its central line has an arbitrary shape but so that its
curvature is weak and its deformation is slow. The section of its core stays
 circular with a little radius $r_0$.

One introduces the local coordinates system $(\tau^A,\rho^a)$ attached to the
central line of the vortex:

\be
\label{5.1}
x^\mu=X^\mu (\tau^A)+\rho^a N^\mu_a (\tau^A)
\ee
The Minkowskian coordinates $x^\mu$ are expressed in function of the two coordinates $\tau^A =(\tau^0 , \tau^3)$ of the world sheet sweeped by the central
line, and of two coordinates $\rho^a=(\rho^1 , \rho^2)$ pointing in a
direction orthogonal to the world sheet along the two orthonormal vectors
$N_a^\mu$. We introduce also polar coordinates:
\be
\label{5.2}
\rho^1=r\cos\phi \quad ,\quad \rho^2=r\sin\phi.
\ee

In this problem we need to define a reference length, choosed unity by
convenience, in order to characterize the spacetime neighbourhood of a
point of the central line in which we shall study the vortex. We shall suppose
that the characteristic length $l$ on which we have a change in the 
tangent plane of the world sheet is large and that the radius of the core
 $r_0$ is small
$$l\gg 1\quad , \quad r_0\ll 1 $$
Let us note that we have two typical lenght $l$ and $r_0$.

The vetors $N_a^\mu$ perpendicular to the world sheet and the tangent vectors
$\frac{\partial X^\mu}{\partial \tau^A}$ have little change on a length unity. We can express
 this fact by
\be
\label{5.5}
N_a^\mu=N_a^\mu (\frac{\tau^A}{l})\quad ,\quad \frac{\partial X^\mu}{\partial \tau^A}=X_{,A}^\mu (\frac{\tau^A}{l})
\ee
hence
\be
\label{5.6}
\partial _A N_a^\mu=\frac{1}{l}N^\mu_{a,A} \quad , \quad \partial _B X^\mu_{,A}=\frac{1}{l}X^\mu_{,AB}.
\ee

The Minkowkian metric $g_{\alpha\beta}$ in the system of coordinates
$(\tau^A,\rho^a)$
can be expressed as
\be
\label{5.7}
g_{AB}=\gamma_{AB}+2K_{aAB}\rho^a+\left(K_{bA}^D K_{aBD}+\delta_{ab}\omega_A\omega_B\right)\rho^a\rho^b
\ee
\be
\label{5.8}
g_{Ab}=\rho^a\epsilon_{ab}\omega_A
\ee
\be
\label{5.9}
g_{ab}=\delta_{ab}
\ee
wherein we recognise the induced metric of the world sheet
\be
\label{5.10}
\gamma_{AB}=X_{,A}^\mu X_{,B}^\nu\eta_{\mu\nu}=O(\frac{1}{l^0}),
\ee
the extrinsic curvature
\be
\label{5.11}
K_{aAB}=\frac{1}{l}\eta_{\mu\nu}N^\mu_{a,A}X^\nu_{,B}=O(\frac{1}{l})
\ee
and the twist defined by
\be
\label{5.12}
\epsilon_{ab}\omega_A=\frac{1}{l}\eta_{\mu\nu}N^\mu_{a,A}N^\nu_b=O(\frac{1}{l}).
\ee
Some other quantities will be useful:
\be
\label{5.13}
\gamma=\det\gamma_{AB},
\ee
the mean curvature
\be
\label{5.14}
K_a=K_{aAB}\gamma^{AB},
\ee
\be
\label{5.15}
g=\det g_{\alpha\beta}=\gamma D^2
\ee
where
\be
\label{5.16}
D=1+K_a\rho^a+\frac{1}{2}(K_aK_b-K^A_{aB}K^B_{bA})\rho^a\rho^b,
\ee
and
\be
\label{5.17}
g^{AB}=\gamma^{AB}-2K_a^{AB}\rho^a+O(\frac{r_0^2}{l^2})
\ee
In the following we suppose that there is no twist in the deformation so that
the cross terms of the metric cancel.

It is natural to adopt the assumption that in the  neighbourhood 
of the core whose the linear dimentions are some unities of $r_0$, the 
components $C_\mu$, expressed in a orthonormal tetrad coincides with the 
solution of the stationary cylindrical vortex described in the preceding 
section, that is with (\ref{26}) which in Minkowskian coordinates, is 
rewritten 
\be
\label{5.18}
C_\nu=(-E , L , -\frac{y}{r^2}M , \frac{x}{r^2}M ),
\ee

We can choose the coordinates $\tau^A$ on the world sheet so that 
$\gamma_{AB}$ takes the conformally flat form:
\be
\label{5.20}
\gamma_{AB}=F(\frac{\tau^A}{l})\eta_{AB}
\ee
Let us note that $\partial _{A}F=O(\frac{1}{l})$.
The holonomic base $\partial_A$, $\partial_a$ is orthogonal but not
orthonormal so that the identification of the solution with the stationary
cylindrical solution must take into account the norm $\sqrt{F}$ of 
$\partial_A$, hence
\be
\label{5.22}
C_\sigma=(-\sqrt{F}E , \sqrt{F}L , -\frac{\rho^2}{r^2}M , \frac{\rho^1}{r^2}M),
\ee
is the solution expressed in the holonomic base $(\partial_A , \partial_a)$.
When expressed in the orthonormal tetrad $(\frac{\partial_A}{\sqrt{F}} , \partial_a)$ we can see that this solution coincides with the solution (\ref{5.18})  of the cylindrical vortex. 

We shall write the equations of motion (\ref{19}) and (\ref{20})  on the 
frontier of the core $r=r_0$and
expand in power of the small parameter $l^{-1}$. The zero
order which is the solution of the straight vortex vanish identically. In this 
expansion appears also the second small parameter $r_0$ .

For this purpose it will be useful to rewrite the expression (\ref{5.22})
 in a more condensate form:
\be
\label{5.25}
C_\sigma=(\overline{C}_A\sqrt{F}\, , \, C_a),
\ee
where
\be
\label{5.27}
\overline{C}_A=(-E , L)\quad ,\quad C_a=( -\frac{\rho^2}{r^2}M\, , \, \frac{\rho^1}{r^2}M) 
\ee
$C_a$ is of order $r_0^{-1}$ .

We must calculate some quantities. Using (\ref{5.17}), (\ref{5.20}), the 
contravariant components $C^\alpha$ are given by
\be
\label{5.29}
C^A=g^{AB}C_B=\frac{1}{\sqrt{F}}\eta^{AB}\overline{C}_B-2K_a^{AB}\rho^a \sqrt{F}\overline{C}_B+\ldots
\ee
\be
\label{5.30}
C^a=\delta^{ab}C_b.
\ee
In (\ref{5.29}), the second term is of order $\frac{r_0}{l}$. The
ellipse designates smaller terms; we shall follow this convention
below. We can express
\be
\label{5.31}
h^2=-g^{\alpha\beta}C_\alpha C_\beta=-g^{AB}C_A C_B-\frac{M^2}{r^2}
\ee
in function of the corresponding quantity of the straight
vortex which will be renamed
\be
\label{5.32}
h^2_0=-\eta^{AB}\overline{C}_A\overline{C}_B-\frac{M^2}{r^2}.
\ee
We obtain
\be
\label{5.33}
h^2=h^2_0+2K_a^{AB}\rho^a\overline{C}_A\overline{C}_{B}F+\ldots
\ee
and
\be
\label{5.34}
W (h^2)=W (h_0^2)+2K_a^{AB}\rho^a\overline{C}_A\overline{C}_{B}F\frac{\partial W}{\partial h^2} (h_0^2)+\ldots.
\ee
where we put $W (h^2)=\frac{n(h)}{h}$.

We can now expand the equations of motion on the frontier $r=r_0$ of the core
of the vortex. The  equation (\ref{20}) in coordinates
$(\tau^A , \rho^a)$
\be
\label{5.43}
\frac{1}{FD}\partial_\sigma\left(FDW(h^2) C^\sigma\right)=0
\ee
gives
\be
\label{5.43'}
\frac{1}{FD}\left(W (h_0^2)\eta^{AB}\partial _{A}\sqrt{F}+
W (h_0^2)FK_a C_b\delta^{ab}+\frac{\partial W}{\partial h^2}(h_0^2) F^2\overline{K}_a C_b\delta^{ab}\right)+\ldots=0
\ee
with
\be
\label{5.44}
\overline{K}_a=2K_a^{CD}\overline{C}_C\overline{C}_D
\ee
In this derivation we have used that for any function $G(r)$:
$$\partial_a\left(G(r)\delta^{ab}C_b\right)=0$$
In the equation (\ref{5.43'}) the first term is of order $\frac{1}{l}$, the second and the third are in $\frac{1}{lr_0}$. So we shall 
retain only the second and third terms in $\frac{1}{lr_0}$ which are the 
leading terms of the expantion (\ref{5.43'}):
\be
\label{5.46'}
\left(W (h^2_0)FK_a+\frac{\partial W}{\partial h^2}(h^2_0)F^2 \overline{K}_a\right)C^a=0
\ee
It is easy to see that the leading
term of the equation (\ref{19}) is of order $\frac{1}{l}$. So the 
equation (\ref{19}) does not contribute at the same order as  the 
equation (\ref{20}) and can be forgotten. 
 
The equation (\ref{5.46'}) is written on the frontier $r=r_0$ of 
the core; $C^a$ depends
of the polar angle $\phi$. Since the equation (\ref{5.46'}) must be
verified for all the polar angles when we turn around the core, we obtain 
finally the equation:
\be
\label{5.47'}
W (h^2_0)FK_a+\frac{\partial W}{\partial h^2}(h^2_0)F^2 \overline{K}_a=0
\ee
Since the radius $r_0$ of the core is supposed small, this equation
gives a good approximation of the equation of motion of the central line.
We shall discuss more extensively this equation in the next section. 
It is convenient now to examine the particular case of the stiff material 
where $W(h^2)=\frac{n}{h}=k$. We have $\frac{\partial W}{\partial h^2}=0$ and
the equation (\ref{5.47'}) is reduced to 
\be
\label{5.50}
K_a=0
\ee
which express the Nambu Goto dynamics of the central line in the intrinsic
coordinates $(\rho^a , \tau^A)$.  

Let us project the extrinsic curvature $K_{aAB}$ in the Minkowski space:
\be
\label{5.53}
K^\mu_{AB}=-\delta^{ab}K_{aAB}N^\mu_b=\partial_A\partial_BX^\sigma\left(\delta^\mu_\sigma-\eta_{\rho\sigma}\eta^{CD}e^\rho_Ce^\mu_D\right)
\ee
whith
\be
\label{5.54}
e^\rho_C=\frac{X^\rho_{,C}}{\sqrt{F}}.
\ee
From (\ref{5.50}) and (\ref{5.53}) the calculus of $K^\mu_{AB}\gamma^{AB}$
gives the usual string equations of motion in Minkowski coordinates:
\be
\label{5.54'}
\frac{1}{\sqrt{-\gamma}}\partial _A(\sqrt{-\gamma}\gamma^{AB}X^{\mu}_{,A})=0
\ee

\section{Discussion}

We cast  the equations of motion (\ref{5.47'}) in the interesting form:
\be
\label{5.51}
K_{aAB}S^{AB}=0
\ee
where
\be
\label{5.52}
S^{AB}=2\frac{\partial W}{\partial h^2} (h^2_0)\eta^{AC}\overline{C}_C
\eta^{BD}\overline{C}_D+W (h^2_0)\eta^{AB}
\ee

For a weak deformation of the cylindrical symmetry ($l\gg1$), a rough  
approximation of (\ref{5.54}) gives:
\be
\label{5.55}
e^x_C \sim e^y_C \sim 0\quad ,\quad e^t_0 \sim e^z_3 \sim 1
\ee
So, from (\ref{5.53}) the leading terms are tranverse ($i=x,y$):
\be
\label{5.56}
K^i_{AB} \approx \partial_A\partial_BX^i,
\ee
and the equations of motion of the core are approximated by
\be
\label{5.57}
S^{AB}\partial_A\partial_BX^i=0.
\ee

The equation (\ref{22}) can be written
\be
\label{5.48}
\frac{dW}{dh^2}=\left(\frac{1}{v^2}-1\right)\frac{W}{2h^2},
\ee
If we assume that the fluid has a polytropic equation of state, then 
we have seen at the end of the second section that when $r$ tends towards the 
radius of the core $r_0$, then $h$, resp. $n(h)$, tends towards $m$, resp. 
zero. As a result, the quantities $W$ and $\frac{dW}{dh^2}$ tends towards 
zero but the quotient of $\frac{dW}{dh^2}$ by $W$ tends towards infinity. 
Therefore in the vicinity of $r=r_0$ we can neglect the second term in 
(\ref{5.52}) and after simplification the equation (\ref{5.57}) is written
\be
\label{5.58'}
E^2\partial_0^2X^i+2EL\partial_0\partial_3X^i+L^2\partial_3^2X^i=0
\ee  
This equation is parabolic and cannot represent propagating waves along the 
core of the vortex. To be more specific, we have seen 
in the second section that it is possible by a Lorentz transformation along 
$z$, to choose $L=0$. In this case the equation (\ref{5.58'}) becomes particularly simple. Its solution $$X= a(\tau^3)\tau^0+b(\tau^3)$$ moves off the 
equilibrium linearly with the time. It can represent the motion of the vortex only at the beginning during a short time, since the method of resolution
supposes a small deviation from the initial equilibrium, which is not the case 
at later time.
 
The equation (\ref{5.54'}) can be 
expressed in the preceeding approximation to compare with (\ref{5.58'}):
\be
\label{last}
\partial_0^2X^i-\partial_3^2X^i=0.
\ee
A small perturbation can propagate along $z$.
This result singles out the dynamics of Nambu Goto obtained in stiff 
material.

\section{conclusion}

If we refer to the usual experience in hydrodynamics, it can appear somewhat 
strange that, in a polytropic fluid, a small deformation of the cylindridal 
core of the vortex cannot propagate as along an elastic string. May be 
the model of isentropic irrotational fluid is too restictive. We must 
also remember that in non relativistic fluid, if 
we bend a cylindrical line vortex, we induce infinite quantities which 
are dynamically embarassing \cite{saff}.

On the other hand in stiff material the core of the vortex has the dynamics 
of the Nambu Goto string and consequently small deformations can propagate. 

To conclude, at our knowledge, it is the first time that the Nambu Goto 
dynamics is directly connected to the classical relativistic hydrodynamics.

\end{document}